\documentclass[12pt]{iopart}

\usepackage{graphicx}
\usepackage{xcolor}
\usepackage{siunitx}
\usepackage{wrapfig}
\usepackage{caption}
\usepackage{subcaption}

\usepackage{tikz}
\usetikzlibrary{positioning}
\usepackage{xcolor}

\begin{document}

\title[]{Teaching a plasma physics and engineering class at a liberal arts college using Course-Based Undergraduate Research Experiences (CUREs)}

\author{Bhavesh Ramkorun}

\address{Department of Physics, Sewanee: The University of the South, Sewanee, TN, 37383, US}

\ead{bhramkor@sewanee.edu}

\vspace{10pt}
\begin{indented}
\item[]\today
\end{indented}

\begin{abstract}
Undergraduate research experiences provide students with an opportunity to apply knowledge learned in the classroom to authentic scientific problems. In this work, we describe the development of a special-topics course in Plasma Physics and Engineering (PPE) at a small liberal arts college using a Course-Based Undergraduate Research Experiences (CUREs). During the first half of the semester, students learned fundamental plasma physics and dusty plasma concepts through traditional lectures and homework. During the second half, they applied this knowledge to investigate an unanswered research question: why does the levitation time of growing nanoparticles decrease in the presence of a magnetic field? Students learned and used Langmuir probe and optical emission spectroscopy diagnostics to characterize an argon plasma over a range of magnetic-field strengths. Their measurements showed a general decrease in electron density with increasing magnetic field strength, while optical emission measurements confirmed a reduction in carbonaceous nanoparticle growth-cycle time from acetylene. Using the measured plasma parameters, students estimated the nanoparticle charge and developed the hypothesis that a reduction in electron density leads to a smaller negative dust charge and consequently a weaker electric force available to levitate the particles as a function of increasing magnetic field strength between 50 and 330 Gauss. Students subsequently disseminated their results through research poster presentations at conferences. Course evaluations and E-CLASS responses also indicated positive student experiences with the laboratory and research components of the course. This work demonstrates one approach for integrating plasma physics education and authentic research in an undergraduate curriculum at a small liberal arts institution.
\end{abstract}

\vspace{2pc}


\section{Introduction}

\begin{figure}[ht]
    \centering
    \includegraphics[width=0.8\linewidth]{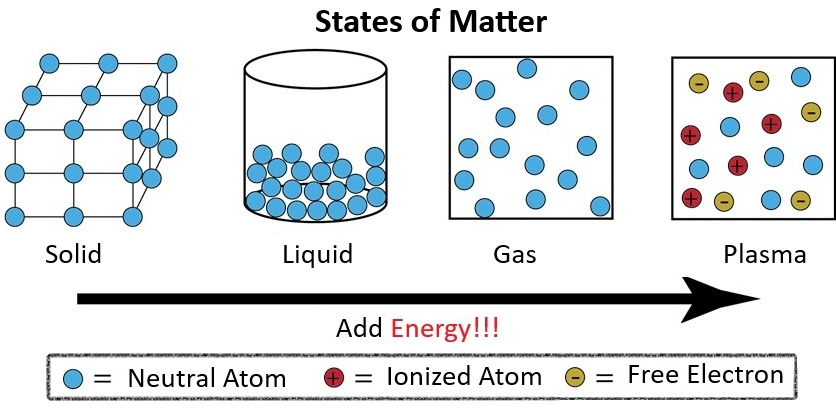}
    \caption{The four states of matter (schematics not drawn to scale): solid, liquid, gas, and plasma.   
     When energy is added (for example: heat), a solid melts into a liquid, the liquid vaporizes into a gas, and the gas ionizes into plasma.}
    \label{fig:4_states_of_matter}
\end{figure}

Plasma is usually considered as the fourth state of matter, with the other three common states being solid, liquid, and gas. Matter is made up of atoms. In solids, the atoms are rigidly connected. In liquids, the atoms are less rigidly connected, free to move and disconnect. In gas, the atoms are disconnected. Plasma occurs when some atoms in gas ionize, that is they lose one electron resulting in free electrons. Plasma is a mixture of gas, ionized atoms and free electrons \cite{adamovich20172017, adamovich20222022}. Energy, for example heat, is needed to convert states of matter from solid to liquid to gas and to plasma. Figure \ref{fig:4_states_of_matter} shows a schematic representation (not to scale) of the four states of matter. 

Plasma is basically an ionized gas, but most importantly, it has its own unique properties \cite{gurnett2005introduction, chen2012introduction, goldston2020introduction}. For example, the separation between ions and electrons gives rise to an intrinsic electric field within the plasma, usually referred to as ambipolar electric field. Several atoms contain excited electrons which upon return to a lower energy level, emit photons at unique wavelengths depending on the gases present in the plasma. These can be measured via optical emission spectroscopy (OES) of the plasma. Furthermore, the individual charged species may respond to a background magnetic field through the Lorentz force. 

Plasma is the dominant state of matter in the universe, but few undergraduate programs provide exposure to plasma science. The Small College Plasma Consortium (SCPC) is ``a coalition of faculty, students, and plasma researchers" primarily involved in undergraduate education \cite{SCPC_about}. According to the SCPC, as of September 2025, there were 29 member institutions involved. The goal of this paper is to show how we developped special topic class in Plasma Physics and Engineering (PPE) at The University of the South, a small liberal arts college and to engage students in ongoing plasma physics research. Since plasma is prominent in our lives, we wish to inspire other faculties at small colleges to create a similar course for teaching plasma physics to undergraduate students.

The course used Course-Based Undergraduate Research Experiences (CUREs) \cite{dolan2016course}, a technique of ``inquiry-based" course development, where undergraduate students must apply course contents to critically think about an unsolved scientific problem. The first half of the semester was spent on learning traditional textbook knowledge about plasma physics. There was weekly homework assignments and a midterm exam for learning and assessment. During the second half of the semester, students worked together on one unanswered research question from plasma physics research, which was ``why do nanoparticles levitate for less time in a magnetized plasma \cite{couedel2019influence, ramkorun2024comparing, ramkorun2025investigating}"? Students read relevant literature from which they formulated hypotheses for the unanswered question. For example, it is likely that the electric force on dusty plasma has decreased during the presence of a magnetic field; hence the particles levitate for less time due to a smaller force balance \cite{tadsen2018probing, ramkorun2025electron}. Students also collected data in the laboratory related to particle growth and levitation to test their hypothesis. 

Students from a wide range of scientific and engineering backgrounds were encouraged to register for the course due to the widespread relevance of plasma physics in key technological areas. For example, plasma physics is applicable in biomedical research to treat bacteria and cancerous cells \cite{laroussi2008plasma, park2012atmospheric, laroussi2015low, keidar2015plasma, gay2016clinical, park2017plasma, karthik2023low}. Plasma is also used in materials science and engineering to grow semiconductor thin films for a wide range of electronic applications \cite{cote1999plasma, park2012review, abliz2020effects, ramkorun2021effects, kaloyeros2023silicon}. Plasma is also used in agricultural science to treat food products such as meat and seeds \cite{jo2014non, colejo2018evaluation, waskow2021mechanisms, zhu2023evaluating}. Finally, plasma is studied in a wide range of areas relevant to physics areas such as fusion \cite{troyon1984mhd, toi2004mhd, tikhonchuk2018physics, frenje2020nuclear} and space physics \cite{hundhausen1970composition, treumann1997advanced, koepke2008interrelated, tsurutani2022space}. 


\section{Experimental set up}

\begin{wrapfigure}{r}{0.48\textwidth}
    \centering
    \includegraphics[width=\linewidth]{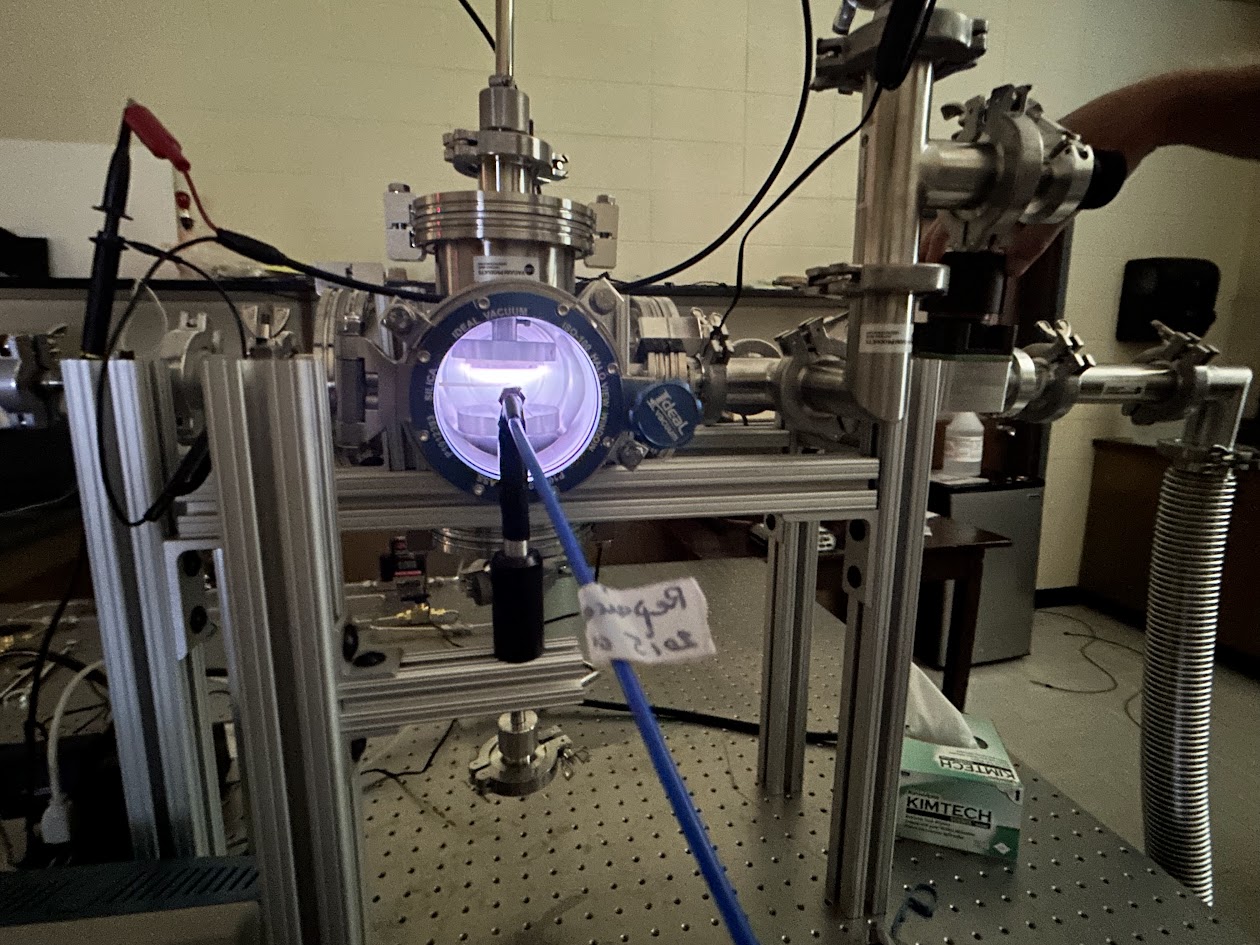}
    \caption{A picture of the experimental chamber used during the CUREs class. It is a capacitively coupled argon plasma that can grow nanoparticles from reactive chemicals. We provided a background magnetic field by using permanent magnets to influence experiments.}
    \label{fig:experimental_chamber}
\end{wrapfigure}

The particle growth chamber was a modified 6-way cross ISO 100 where two ports were used to fit in parallel plate electrodes, three ports were used to fit in window to view the plasma and one port used to flow gases. We used argon (90\%) and acetylene (10\%) gases to turn on the plasma. We used a radiofrequency power supply (13.56 MHz) to power part of the top electrode. The rest of the top electrode was grounded and shieleded from the powered part using an insulating ceramic ring. The bottom electrode was grounded. We also used titanium tetraisopropoxide (TTIP) as a precursor to grow titania nanoparticles and it simply replaced the acetylene line. The TTIP line was heated to 70 $^\circ$ when used. A figure of the plasma between the parallel plate electrodes is shown in Figure \ref{fig:experimental_chamber}.

Students were involved to set up the growth experiment. A few class periods were spent in the laboratory where students gained hands-on experience connecting vacuum lines, installing components such as vacuum pumps and pressure gauge, and testing the gas flow, designing the electrodes, assembling the machined parts and using the power supply to ignite plasma. 


\section{The CUREs research question}

\begin{figure}[ht]
    \centering
    \includegraphics[width=\linewidth]{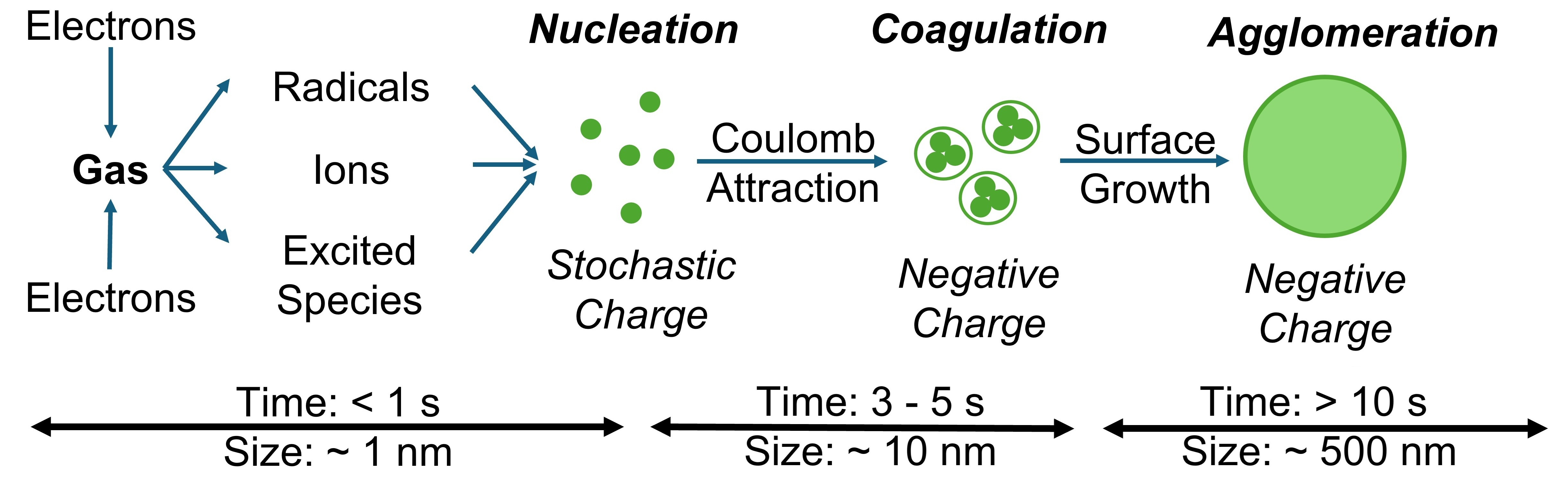}
    \caption{Schematic representation of one growth cycle. Electrons from the backtround plasma bombard chemical gases to produce ions, radicals, and excited species. These can nucleate into small NPs $(\sim 1 nm)$ which can then coagulate into larger particles $(\sim 10 nm)$. Finally, these can agglomerate into bigger particles $(\sim 500 nm)$ through surface growth.}
    \label{fig:figure_growth_steps}
\end{figure}

The research question for this CUREs class is motivated from nanoparticles’ levitation in a dusty plasma. These particles spontaneously nucleate from gaseous chemical precursors in a low-temperature plasma in three steps: nucleation, coagulation, and agglomeration, as schematically shown in Figure \ref{fig:figure_growth_steps} \cite{kortshagen2016nonthermal}. The energetic free electrons from the background plasma bombard the gas and dissociate them into radicals, ions and excited species. These chemically react to each other and nucleation into nanoparticles that are about 1 nm in dimensions and stochastically charged \cite{ravi2009coagulation}. Next, they coagulate, that is they can attract to each other and stick together. They grows them to about 10 nm and they are now negatively charged from electrons in the background plasma. Now, they all repel each other and we move into the agglomeration phase where surface growth occurs due to reaction with the background neutrals.

As they grow, they remain levitated in the plasma for a period of time known as the growth cycle, after which they leave the plasma \cite{donders2022characterization, ramkorun2024introducing}. Their levitation results from a balance of forces, including electric, gravitational, and ion-drag forces \cite{van2015fast}. When a background magnetic field is applied during nanoparticles' synthesis, the duration of the growth cycle decreases \cite{ramkorun2025electron, ramkorun2024comparing}. This leads to the central research question: why does the growth cycle shorten in the presence of a magnetic field? The end-goal of the class was for students to perform some experiments to understand the changes in the properties of a magnetized plasma and to formulate a hypothesis towards the research question.

We reviewed several literature as a class to understand the scientific evidence. We saw at least four different studies spanning 15 years that showed evidence of a reduced cycle time for carbonaceous particle synthesis in an admixture of argon and acetylene in a capacitively coupled plasma while exposed to a background magnetic field. First, Greiner et. al (2011) showed a reduced cycle time from 180 seconds (s) at 33 milliTesla (mT) (330 Gauss) to 120 s at 67 mT \cite{greiner2011nanodust}. Second, Couedel et. al (2019) showed a reduced cyce time from about 40 to 60 s without a background magnetic field down to 15-20 s with a background magnetic field of up to 32 milliTesla \cite{couedel2019influence}. Third, Ramkorun et. al (2024) showed a reduced growth cycle from about 120 s without magnetic field to about 40 s with a background magnetic field of 500 Gauss \cite{ramkorun2024comparing}. Finally, Ramkorun et. al (2025) showed a gradual decrease in the cycle time from 120 s without a background magnetic field to 40 s at the onset of the electron magnetization after which the cycle time remained approximately constant \cite{ramkorun2025electron}. From a combination these papers, it was obvious that the experimental results were reproducible and independent of researchers. The next step was for the students to built a foundational learning of plasma physics in order to measure and analyze plasma properties relevant to the research question. 


\section{The Course Content}

\begin{figure}[t]
    \centering
    \begin{subfigure}[t]{0.48\linewidth}
        \centering
        \includegraphics[height=4.25cm]{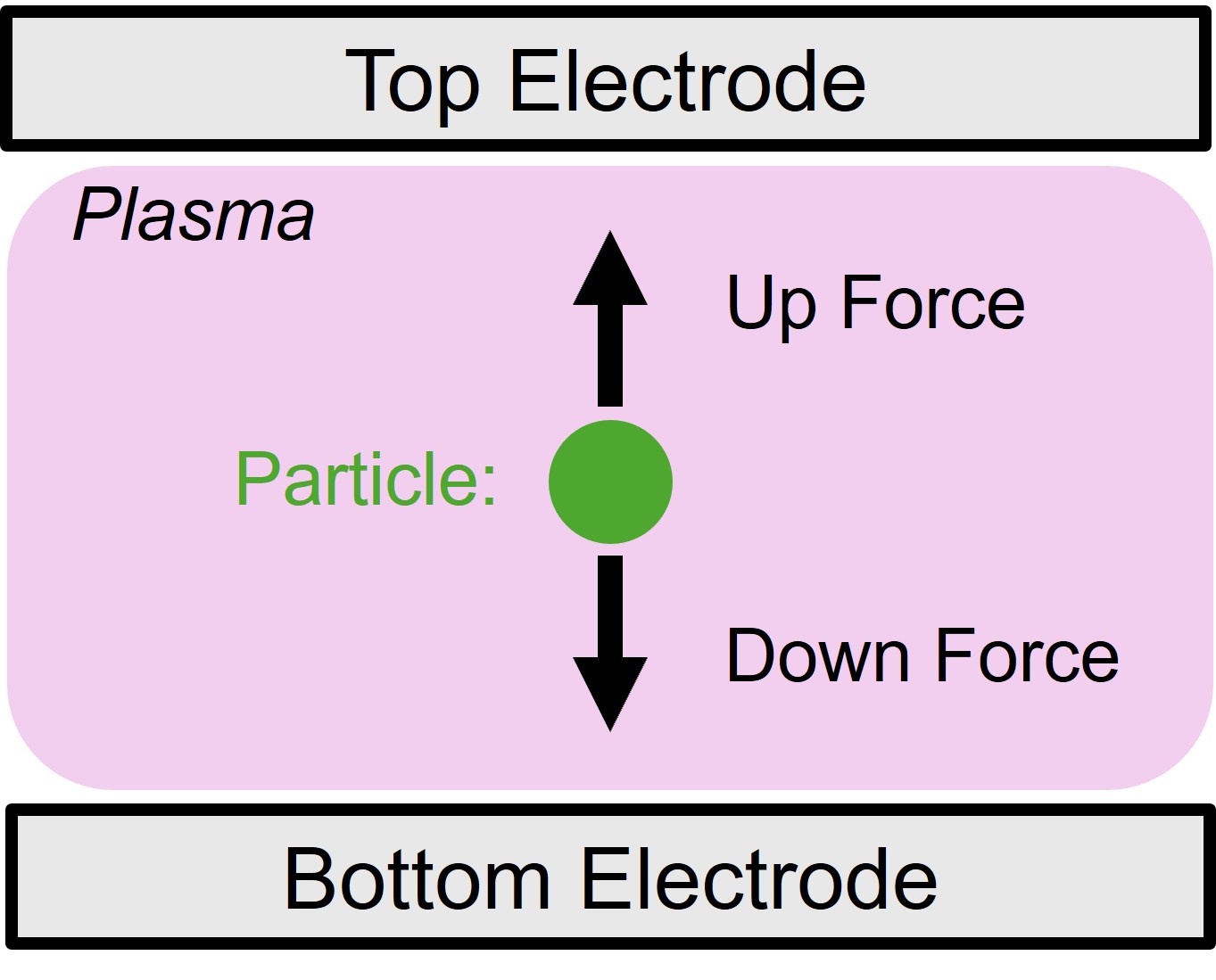}
        \caption{}
        \label{fig:force_balance}
    \end{subfigure}
    \hfill
    \begin{subfigure}[t]{0.48\linewidth}
        \centering
        \includegraphics[height=4.15cm]{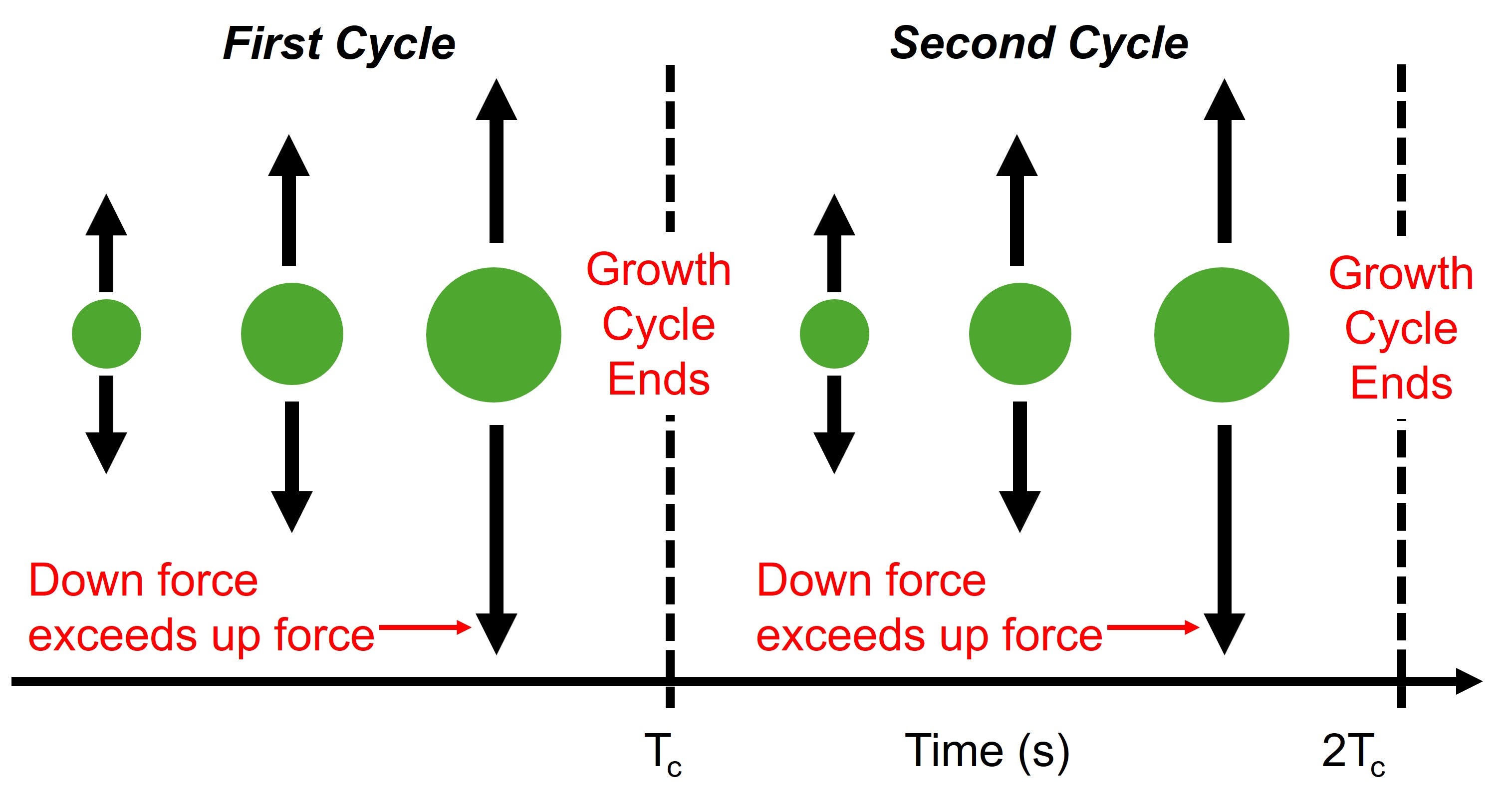}
        \caption{}
        \label{fig:growth_cycle}
    \end{subfigure}      

    \caption{A schematic representation of the force balance for nanoparticles in a dusty plasma. (a) An ``up force" balancing a ``down force" to keep a nanoparticle levitated during growth between the electrodes in the plasma, and (b) as the particle grows bigger, the downward force exceeds the upward force, which breaks the force balance, and the growth cycle ends. }
    \label{fig:schematic_force_balance_and_cycle}
\end{figure}

The course content was organized so that students learned a major topic every week before the midterm of the semester. They directly engaged with the CUREs research question during the second half of the semester by performing experiments, reading literature reviews and making a poster presentation. The course began by introducing dusty plasmas, with emphasis on dust charging and the gravitational, electric, neutral-drag, thermophoretic, and ion-drag forces that influence particle levitation. A schematic description of the levitation within a particle growth cycle is shown in figure \ref{fig:schematic_force_balance_and_cycle}.

In the following weeks, they learned about the fundamental properties of low temperature plasmas, 
using single-particle motion. The topics included degree of ionization, Poisson equation and electric potential, plasma frequency, Debye length, quasi-neutrality, and the response of charged particles to electric ${\bf E}$ and magnetic fields ${\bf B}$ including cyclotron motion, gyroradius, and the ${\bf E}\times{\bf B}$ drift. We also spent time analyzing the confinement forces needed to levitate particles in dusty plasma such as electric, gravitation, ion-drag, neutral drag and thermophoretic forces.  There were weekly homework on each topics and at the middle of the semester, there was a comprehensive midterm exam.

After the midterm exam and semester break, students learned about plasma diagnostics that they would subsequently use in the laboratory component of the CURE. We spent one week on OES, beginning with atomic excitation and the Boltzmann distribution and progressing to spectral-line intensities, line broadening, and the use of Boltzmann plots to estimate excitation temperature. Then, we spent a week on Langmuir probes, where we developed the underlying theory of ion and electron saturation currents, and floating potential, and analyzed current--voltage ($I$--$V$) characteristics to determine electron temperature, electron density, and plasma potential. These provided a transition from theoretical instruction to the research component of the course. For the rest of the semester, students applied these diagnostic techniques to laboratory measurements, including experiments performed at varying magnetic-field strengths, before moving into literature discussion, analysis of their results, and preparation of their final research posters.


\section{Implementing CURE}

The students measured the background argon plasma properties using either OES or a flat-disk Langmuir probe. The goal of the OES was to have a non-invasive diagnostics during particle synthesis for us to experimentally confirm there was a reduced levitation time as a function of magnetic field strength. The goal of the probe was to detect any changes in the electron temperature or density of the background plasma as a function of magnetic fields before particle growth. 

\subsection{Langmuir probes}

\begin{figure}[htbp]
    \centering

    \begin{subfigure}[t]{\textwidth}
        \centering
        \includegraphics[width=\linewidth]{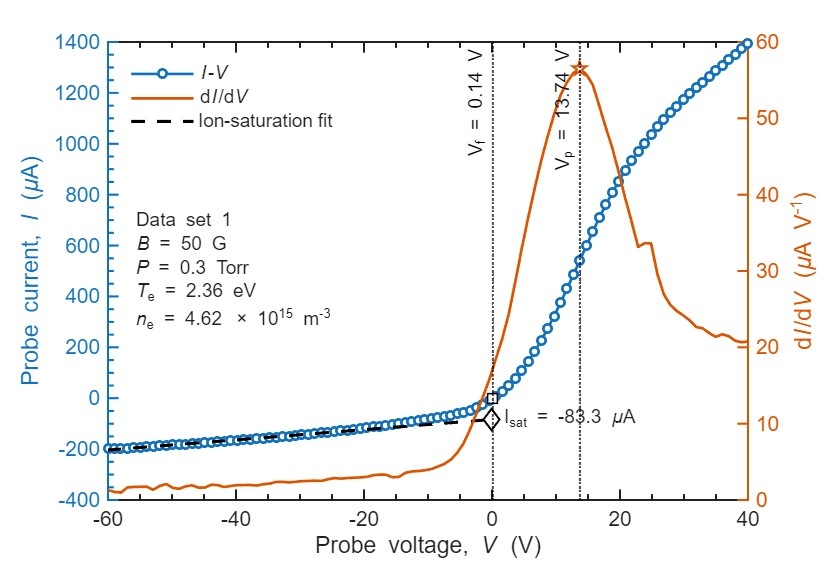}
        \caption{}
        \label{fig:langmuir_iv}
    \end{subfigure}


    \begin{subfigure}[t]{0.487\textwidth}
        \centering
        \includegraphics[width=\linewidth]{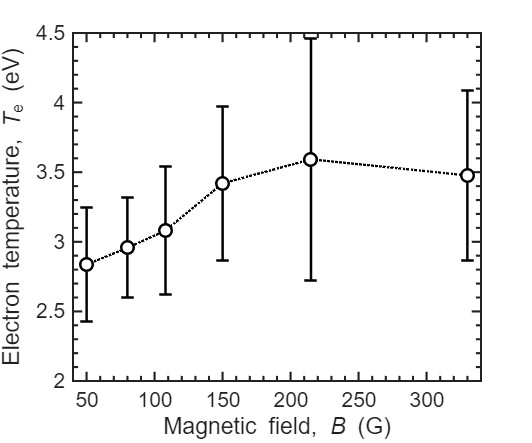}
        \caption{}
        \label{fig:Te_vs_B}
    \end{subfigure}
    \hfill
    \begin{subfigure}[t]{0.487\textwidth}
        \centering
        \includegraphics[width=\linewidth]{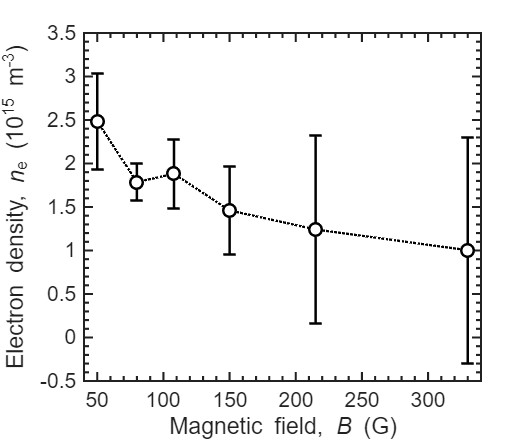}
        \caption{}
        \label{fig:ne_vs_B}
    \end{subfigure}
    \caption{%
    Langmuir probe characterization of the argon plasma at a pressure of
    0.3 Torr. (a) An example of a current-voltage (I-V) plot (primary axis) and
    corresponding first derivative of current with respect to voltage (dI/dv) (secondary y-axis) used to determine the plasma parameters.
    (b) Electron temperature, $T_e$, as a function of magnetic field strength.
    (c) Electron density, $n_e$, as a function of magnetic field strength.
    Error bars in (b) and (c) arise from one standard deviation of three
    independent measurements.
    }    \label{fig:langmuir_results}

\end{figure}

Using a flat disk Langmuir probe (diameter = 10 mm, thickness = 1 mm), we measured a current-voltage (I-V) trace of the argon plasma at an experimental pressure of 0.3 Torr for a range of magnetic field strength from 50 to 330 Gauss. An example of the I-V trace at 50 Gauss is shown in Figure \ref{fig:langmuir_iv}. Then, by closely following the literature from Robert Merlino (2007), the students were able to extract the ion saturation current, plasma potential (peak in dI/dV as a function of V) from which they calculated the electron temperature $(T_e)$ and electron density $(n_e)$ of the background argon plasma \cite{merlino2007understanding}. The experiments were repeated three times to calculate an average and standard deviation of $T_e$ and $n_e$ for each magnetic field strength which are plotted in figures \ref{fig:Te_vs_B} and \ref{fig:ne_vs_B}, respectively. 

\subsection{Optical Emission Spectroscopy}

\begin{figure}[htbp]
    \centering

    \begin{subfigure}[t]{0.49\linewidth}
        \centering
        \includegraphics[width=\linewidth]{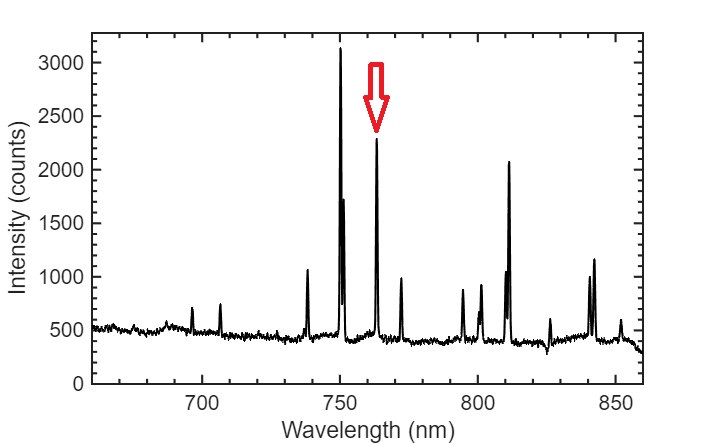}
        \caption{}
        \label{fig:survey_OES}
    \end{subfigure}
    \hfill
    \begin{subfigure}[t]{0.49\linewidth}
        \centering
        \includegraphics[width=\linewidth]{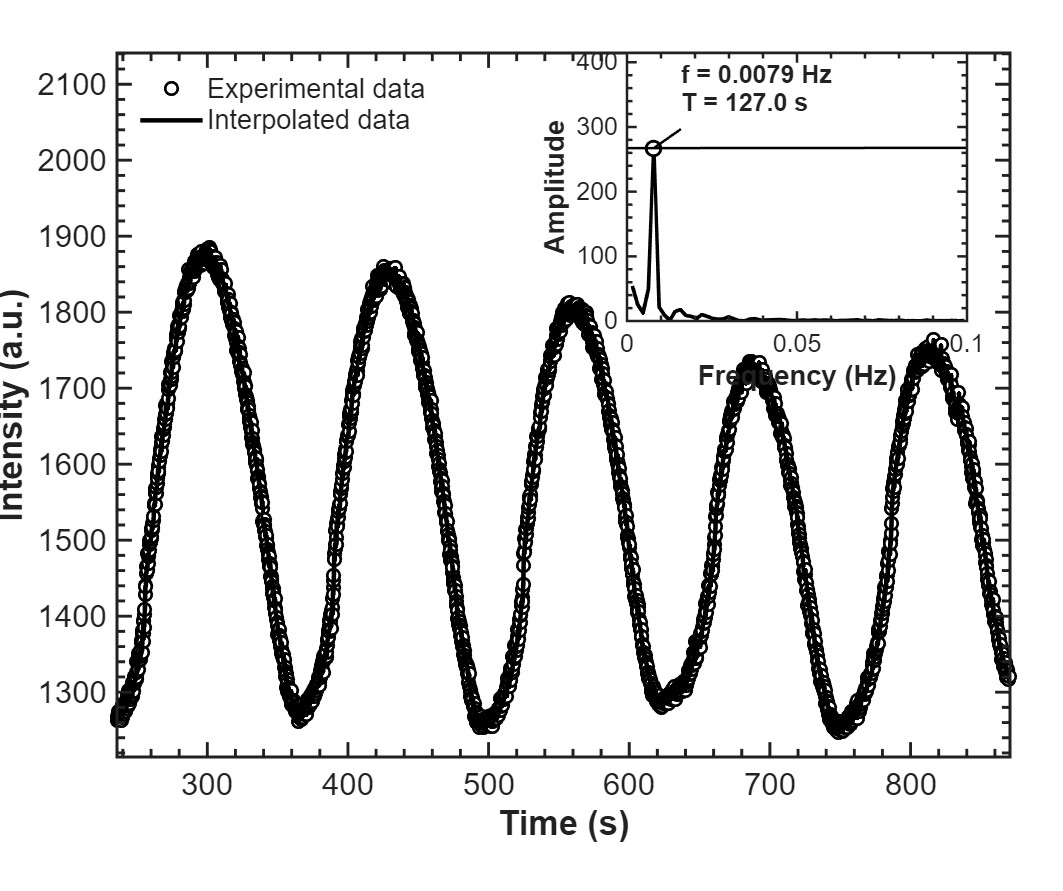}
        \caption{}
        \label{fig:cycle_oes}
    \end{subfigure}

    \vspace{0.5em}

    \begin{subfigure}[t]{0.49\linewidth}
        \centering
        \includegraphics[width=\linewidth]{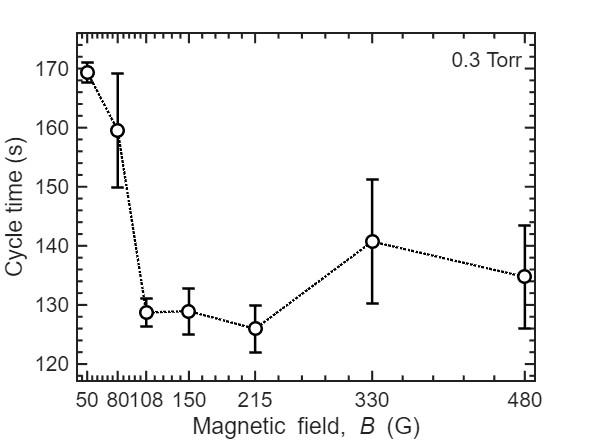}
        \caption{}
        \label{fig:oes_cycle_times}
    \end{subfigure}

    \caption{Optical emission spectroscopy (OES) analysis of the background
    argon plasma during nanoparticle synthesis.
    (a) A survey scan between 650 and 850 nm showing several argon spectral
    lines. The red arrow points towards the intensity at 763.5 nm.
    (b) Temporal variation of the intensity at 763.5 nm, showing a frequency
    of 0.0079 Hz, corresponding to an approximate levitation cycle time of
    127 s.
    (c) The cycle time as a function of magnetic field strength. The data was collected three times and averaged and the standard deviation were set as the error bars.}

    \label{fig:oes_analysis}
\end{figure}

The students used a USB spectrometer to collect a survey scan of the plasma during nanoparticle synthesis. For example, data collected between 650 and 850 nm are shown in figure \ref{fig:survey_OES}. The resolution of our spectrometer (0.56 nm) was not fine enough for students to detect any peaks the chemical gas (acetylene). Furthermore, the ratio of argon to acetylene was 7:1 which means there was only a small trace amount of species from the chemical gas in the plasma. Hence, we only saw peaks related to argon.

All the argon peaks' intensity vary as a function of time during the particle growth to show the growth cycle time, thus corresponding to the levitation time. Here, we only chose one peak at 763.5 nm, because it was relatively isolated (red arrow) in figure \ref{fig:survey_OES} to observe its temporal variation. The intensity variation is plotted in figure \ref{fig:cycle_oes}. We used a fast fourier transform (FFT) to extract the peak frequency, which was the inverse of the cycle time. Since the data were collected at irregular time intervals of about 0.2 to 0.3s, we interpolated them to a constant time step of 0.1 s using spline interpolation before perfoming the FFT. The FFT is in the top-left inset of figure \ref{fig:cycle_oes}. This is one data at 108 Gauss which corresponded to a cycle time of about 127 seconds, however, we repeated the experiments three times to average each cycle time. 

\subsection{Forming a hypothesis from the experimental results}

After the students spent several weeks collecting and analyzing data, they started to develop their hypothesis. Their main source of data to guide their hypothesis were from figures \ref{fig:ne_vs_B} and \ref{fig:oes_cycle_times}. There seemed to be a correlation between a decrease in both electron density and cycle time as a function of increasing magnetic field strength. The students then went back to their notes and the literature to find a relation between the dust charge and the electron density in the plasma, which was equation \ref{eq:dust_potential} \cite{ thomas2010driven}.

\begin{equation}\label{eq:dust_potential}
\left( 1 + \frac{4 \pi \varepsilon_0 r_d n_d \phi_s}{e n_e} \right)
\left( \frac{m_i T_e}{m_e T_i} \right)^{1/2}
\exp\!\left( \frac{e \phi_s}{k_B T_e} \right) \\
= 1 - \frac{e \phi_s}{k_B T_i}
\end{equation}

This equation presents a nonlinear relationship between the dust surface potential $(\phi_s)$ and plasma parameters such as dust density $(n_d)$, and ion temperature $(T_i)$, and $T_e$. $r_d$ is the radii of the dust particles. In a our plasma  we assume can $T_i \approx 300~\mathrm{K}$. When dust is introduced, $N_e$ decreases because electrons stick on the dusts' surface to charge them, while $N_i$ remains approximately constant. Thus, $N_e = N_i$ and $T_e$ can be estimated from a clean argon plasma using a Langmuir probe. The dust density $N_d$ is more difficult to measure, but can be estimated from dust density waves \cite{thomas2010driven, tadsen2018probing, melzer2020dust}, and is typically $\sim 10^{13}~\mathrm{m^{-3}}$ for carbonaceous nanoparticles synthesized from gaseous precursors, similar to our experiment. The dust charge $Q_d$ is then given by equation \ref{eq:dust_charge}, where $\epsilon_o$ is the permittivity of free space.

\begin{equation}
    Q_d = 4 \pi \epsilon_o r_d^2 \phi_s
    \label{eq:dust_charge}
\end{equation}

\begin{figure}[ht]
    \centering
    \includegraphics[width=0.75\linewidth]{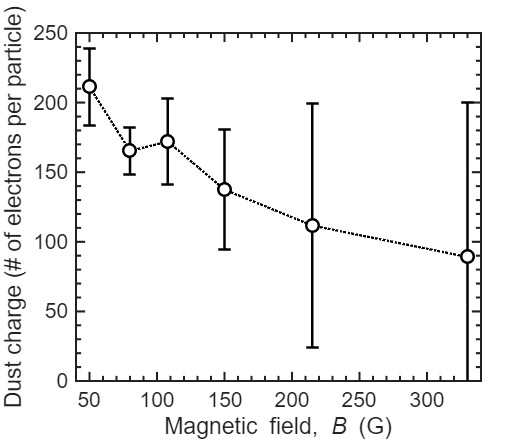}
    \caption{A numerical estimation of the dust charge as a function of magnetic field strength after solving equations \ref{eq:dust_potential} and \ref{eq:dust_charge} and assuming the particles radii to be 100 nm and other plasma parameters from the Langmuir probe results \ref{fig:langmuir_results}. There seems to be a general decrease in dust charge which also coincide with a decrease in cycle time from figure \ref{fig:cycle_oes}, hence students hypothesized that a decease in dust charge is causing the particles to levitate for less time due to a reduced electric force.}
    \label{fig:figure_dust_charge}
\end{figure}

Since students did not have access to information about the dust size during these particle growth experiments, they simply estimated the dust charge for  a constant radius of $r_d = 100$ nm.  They took the measured data for $n_e$ and $T_e$ from the Langmuir probe to numerically solve equation \ref{eq:dust_potential} and calculated the dust charge using equation \ref{eq:dust_charge}. The result is plotted in figure \ref{fig:figure_dust_charge} which also shows a gradual decrease in dust charge with an increasing magnetic field strength. 

It was reasonable for students to question whether a reduced dust charge was even possible. In fact, Tadsen et al. (2018) showed that dust charge can decrease from about 60 electrons down to about 10 electrons with an increasing magnetic field from 0 to 1000 Gauss in a similar plasma containing carbonaceous particles grown from acetylene \cite{tadsen2018probing}. They obtained in-situ information about plasma parameters such as electron and ion density during the actual particle growth experiment, which made their estimation different from ours. However, our numerical calculation did not seem abnormal. 

Since the balance of electric force and gravitational force levitates the particles in the plasma, students hypothesized that a decrease in electron density due to the increased magnetic field strength let to a lower dust charge and lower electric force to levitate the particles, and hence they were levitating for a smaller time in the magnetized plasma.  Several hypotheses are shown in figure \ref{fig:student_hypothesis}.


\begin{figure}[htbp]
\centering

\begin{tikzpicture}[
    studentbox/.style={
        draw=black,
        line width=0.8pt,
        rounded corners=2pt,
        text width=0.90\textwidth,
        inner sep=10pt,
        align=left,
        font=\normalsize
    },
    title/.style={
        font=\bfseries,
        fill=white,
        inner xsep=5pt
    },
    node distance=0.35cm
]

\node[studentbox] (s1) {
``In the presence of a magnetic field, the movement of electrons is restricted
perpendicular to the field lines (by right-hand rule), thereby increasing the
electron density. The stronger electric field thus generates the necessary
$F_E$ to levitate smaller particles.''
};

\node[title, anchor=west] at ([xshift=8pt]s1.north west)
{Student 1};

\node[studentbox, below=of s1] (s2) {
``The smaller levitated particle size in the magnetized plasma can be explained
by a reduction in electron density $n_e$. This reduces the negative dust charge,
which lowers the upward electric force, so only smaller particles remain
levitated.''
};

\node[title, anchor=west] at ([xshift=8pt]s2.north west)
{Student 2};

\node[studentbox, below=of s2] (s3) {
``The magnets could decrease the electron density ($n_e$). This would mean less
net charge and less electric force.''
};

\node[title, anchor=west] at ([xshift=8pt]s3.north west)
{Student 3};

\node[studentbox, below=of s3] (s4) {
``When there is a magnetic field applied to the plasma it reduces the amount of
electrons that move within the field making the electron density smaller. When
there are fewer electrons available this makes the electric force smaller so
the maximum size becomes smaller for a particle to be.''
};

\node[title, anchor=west] at ([xshift=8pt]s4.north west)
{Student 4};

\node[studentbox, below=of s4] (s5) {
``A reduction in electron density decreases the negative charge on dust
particles, weakening the electrostatic force and allowing only smaller
particles to levitate.''
};

\node[title, anchor=west] at ([xshift=8pt]s5.north west)
{Student 5};

\end{tikzpicture}

\caption{The hypothesis submitted by five students in the CUREs class after analyzing the data to explain why particles levitate for smaller amount of time in the plasma; some students inferred a smaller particle levitation time would lead to a smaller particle size after reading Couedel et. al. (2019) \cite{couedel2019influence}, Ramkorun et al. (2024) \cite{ramkorun2024comparing}, Ramkorun et. al (2025) \cite{ramkorun2025electron} who showed a smaller nanoparticle size with increasing magnetic field strength. Unfortunately, we did not have the resources to measure particle size in this CUREs class.}
\label{fig:student_hypothesis}

\end{figure}


\subsection{Scientific Dissemination}

One important aspect of CURE is to disseminate results that students find. Students were encouraged to submit an abstract to conferences based on their formulated hypothesis and experimental results. This led to three poster presentations at IEEE International Conference on Plasma Science (ICOPS) 2026 \cite{villaflor2026detecting, buckner2026optical, hyden2026studying} and two poster presentations at APS Division of Plasma Physics (DPP) 2026. The students who participated in the poster sessions remained active in PPE and STEM after the CUREs class and are currently seeking future opportunities in PPE after graduating from college. 


\section{Class Evaluation}

Six students took the class. The class was evaluated using anonymous university class evaluations at the end of the semester and an anonymous Colorado Learning Attitudes about Science Survey for Experimental Physics (E-CLASS) survey to investigate students' views about experimental physics 
before and after the course \cite{lewandowski2014colorado, wilcox2016students, werth2023enhancing, liu2025students}. The results show that students not only had a positive experience learning PPE but also want to remain involved in physics beyond the classroom.

\subsection{Student Evaluations}

The class evaluation had a 100\% response rate. Student indicated a positive experience with both the course content and the CUREs format. All students strongly agreed that the course content was effectively communicated. Students also reported substantial engagement with  the course outside of class, with five of the six students reporting spending 5--8 hours per week studying and preparing for the course and one student reporting spending 9--12 hours per week. These were times that student spent either doing studying, doing homework, reading research papers, or analyzing laboratory data.

The written responses were particularly useful for evaluating the impact of combining traditional instruction with the CUREs. When 
students were asked what they appreciated most about the course, several specifically identified the laboratory and research component. For example, students wrote ``I enjoyed seeing the process of working in a lab and doing experiments'', ``The lab work and subsequent analyses we did (CURE method basically)", ``I liked that this course was both lecture based and lab based. I like that the first half was lecture based and it really helped me build off of what I had learned so that I could apply it to second half of the semester".

\subsection{E-CLASS Survey}

Five students completed both E-class surveys, which only represented 83\% of the class. Furthermore, due to the small class size, these results should not be generalized beyond this particular group of students. However, the responses provide useful insights into how students' views of experimental physics evolved during the CUREs class.

Four of the five students (80\%) who completed the survey reported that  their ``interest in physics" increased during the semester, while one student (20\%) reported that their interest remained the same. No students reported a  decrease in their interest in physics. 

There was many other positive results from the survey such as students reported an "feel confident to learn new piece of lab equipment", "working in a group is an important part of doing physics experiments", "scientific journals are helpful for answering questions and designing experiments". These kind of attitude is important to success in STEM and the CUREs class provide a real-life example of that in experimental dusty plasmna physics.


\section{Conclusion}

In this work, we developed a special-topics course in Plasma Physics and Engineering using a Course-Based Undergraduate Research Experiences at a small liberal arts college. The course was intentionally divided between traditional classroom instruction and an authentic research experience. During the first half of the semester, students developed the fundamental plasma physics knowledge needed to understand dusty plasma and particle levitation. During the second half, they learned plasma diagnostic techniques and applied them in the laboratory to investigate why nanoparticles levitate for a shorter time in the presence of a magnetic field.

An important outcome of the CUREs was that the students were not simply reproducing a known laboratory experiment. Instead, they collected and analyzed experimental data and used their results together with the literature to develop a hypothesis for an unanswered research question. Langmuir probe measurements showed a general decrease in electron density as the magnetic field increased, while OES measurements showed a reduction in the nanoparticle growth-cycle time. Using the measured plasma parameters, students estimated that the dust charge also decreased with increasing magnetic field strength. From these observations, they hypothesized that the decrease in electron density reduces the negative charge on the nanoparticles and consequently reduces the electric force available to levitate them. This provides a possible explanation for the shorter particle levitation time observed in the magnetized plasma, although additional measurements of the plasma and particle properties during nanoparticle growth will be needed to test this hypothesis directly.

The CUREs also provided students with experience beyond the traditional classroom. They assembled and operated plasma experiments, learned Langmuir probes and optical emission spectroscopy diagnostics, analyzed experimental data, read scientific literature, developed hypotheses, and communicated their results. Several students subsequently presented this work at IEEE ICOPS 2026. Student evaluations showed a positive response to combining lecture-based instruction with laboratory research, and four of the five students who completed both E-CLASS surveys reported an increased interest in physics during the semester. Because the course included only six students, these educational results should not be generalized beyond this particular class, but they provide encouraging evidence for this approach.

Overall, this course shows that plasma physics can provide a useful platform for a CUREs because it naturally combines fundamental physics, engineering, experimental diagnostics, computation, and an active research question. A similar approach could be adapted at other undergraduate institutions to broaden students' exposure to plasma science while providing an authentic research experience within the structure of a course.


\section*{Acknowledgement}

Thank you Dr. Nathan D. Davis (University of Mississippi) for visiting our classroom and providing insightful discussions on CUREs, and undergraduate physics education.  Thank you Dr. Saikat C. Thakur (Auburn University) for visiting our classroom,  providing insightful discussions Langmuir probes and reviewing students' posters.


\section*{References}
\providecommand{\newblock}{}


\begin{thebibliography}{10}
\expandafter\ifx\csname url\endcsname\relax
  \def\url#1{{\tt #1}}\fi
\expandafter\ifx\csname urlprefix\endcsname\relax\def\urlprefix{URL }\fi
\providecommand{\eprint}[2][]{\url{#2}}

\bibitem{adamovich20172017}
Adamovich I, Baalrud S~D, Bogaerts A, Bruggeman P~J, Cappelli M, Colombo V, Czarnetzki U, Ebert U, Eden J~G, Favia P {\em et~al.\/} 2017 {\em Journal of Physics D: Applied Physics\/} {\bf 50} 323001

\bibitem{adamovich20222022}
Adamovich I, Agarwal S, Ahedo E, Alves L~L, Baalrud S, Babaeva N, Bogaerts A, Bourdon A, Bruggeman P~J, Canal C {\em et~al.\/} 2022 {\em Journal of Physics D: Applied Physics\/} {\bf 55} 373001

\bibitem{gurnett2005introduction}
Gurnett D~A and Bhattacharjee A 2005 {\em Introduction to plasma physics: with space and laboratory applications\/} (Cambridge university press)

\bibitem{chen2012introduction}
Chen F~F 2012 {\em Introduction to plasma physics\/} (Springer Science \& Business Media)

\bibitem{goldston2020introduction}
Goldston R~J 2020 {\em Introduction to plasma physics\/} (CRC Press)

\bibitem{SCPC_about}
Small college plasma consortium \url{https://smallcollegeplasma.org/about/} accessed: 2025-10-02

\bibitem{dolan2016course}
Dolan E~L 2016 {\em Natl Res Counc Comm Pap\/} {\bf 1} 1--34

\bibitem{couedel2019influence}
Cou{\"e}del L, Artis D, Khanal M, Pardanaud C, Coussan S, LeBlanc S, Hall T, Thomas~Jr E, Konopka U, Park M {\em et~al.\/} 2019 {\em Plasma Research Express\/} {\bf 1} 015012

\bibitem{ramkorun2024comparing}
Ramkorun B, Chandrasekhar G, Rangari V, Thakur S~C, Comes R~B and Thomas~Jr E 2024 {\em Plasma Sources Science and Technology\/} {\bf 33} 115004

\bibitem{ramkorun2025investigating}
Ramkorun B {\em et~al.\/} 2025

\bibitem{tadsen2018probing}
Tadsen B, Greiner F and Piel A 2018 {\em Phys. Rev. E\/} {\bf 97} 033203

\bibitem{ramkorun2025electron}
Ramkorun B, Thakur S~C, Comes R~B and Thomas~Jr E 2025 {\em Physical Review E\/} {\bf 112} 045211

\bibitem{laroussi2008plasma}
Laroussi M, Hynes W, Akan T, Lu X and Tendero C 2008 {\em IEEE Transactions on plasma science\/} {\bf 36} 1298--1299

\bibitem{park2012atmospheric}
Park G, Park S, Choi M, Koo I, Byun J, Hong J, Sim J, Collins G and Lee J 2012 {\em Plasma Sources Science and Technology\/} {\bf 21} 043001

\bibitem{laroussi2015low}
Laroussi M 2015 {\em IEEE transactions on plasma science\/} {\bf 43} 703--712

\bibitem{keidar2015plasma}
Keidar M 2015 {\em Plasma Sources Science and Technology\/} {\bf 24} 033001

\bibitem{gay2016clinical}
Gay-Mimbrera J, Garc{\'\i}a M~C, Isla-Tejera B, Rodero-Serrano A, Garc{\'\i}a-Nieto A~V and Ruano J 2016 {\em Advances in therapy\/} {\bf 33} 894--909

\bibitem{park2017plasma}
Park J~Y, Park S, Choe W, Yong H~I, Jo C and Kim K 2017 {\em ACS Applied Materials \& Interfaces\/} {\bf 9} 43470--43477

\bibitem{karthik2023low}
Karthik C, Sarngadharan S~C and Thomas V 2023 {\em International Journal of Molecular Sciences\/} {\bf 25} 524

\bibitem{cote1999plasma}
Cote D~R, Nguyen S, Stamper A~K, Armbrust D~S, Tobben D, Conti R~A and Lee G~Y 1999 {\em IBM Journal of research and development\/} {\bf 43} 5--38

\bibitem{park2012review}
Park J~S, Maeng W~J, Kim H~S and Park J~S 2012 {\em Thin solid films\/} {\bf 520} 1679--1693

\bibitem{abliz2020effects}
Abliz A 2020 {\em Journal of Alloys and Compounds\/} {\bf 831} 154694

\bibitem{ramkorun2021effects}
Ramkorun B, Chakrabarty K and Catledge S~A 2021 {\em Materials Research Express\/} {\bf 8} 046401

\bibitem{kaloyeros2023silicon}
Kaloyeros A~E and Arkles B 2023 {\em ECS Journal of Solid State Science and Technology\/} {\bf 12} 103001

\bibitem{jo2014non}
Jo Y~K, Cho J, Tsai T~C, Staack D, Kang M~H, Roh J~H, Shin D~B, Cromwell W and Gross D 2014 {\em Crop Science\/} {\bf 54} 796--803

\bibitem{colejo2018evaluation}
Colejo S, Alvarez-Ord{\'o}{\~n}ez A, Prieto M, Gonz{\'a}lez-Raurich M and L{\'o}pez M 2018 {\em Innovative Food Science \& Emerging Technologies\/} {\bf 50} 84--93

\bibitem{waskow2021mechanisms}
Waskow A, Howling A and Furno I 2021 {\em Frontiers in Physics\/} {\bf 9} 617345

\bibitem{zhu2023evaluating}
Zhu W, Tan G, Han M, Bu Y, Li X and Li J 2023 {\em Innovative Food Science \& Emerging Technologies\/} {\bf 84} 103286

\bibitem{troyon1984mhd}
Troyon F, Gruber R, Saurenmann H, Semenzato S and Succi S 1984 {\em Plasma physics and controlled fusion\/} {\bf 26} 209

\bibitem{toi2004mhd}
Toi K, Ohdachi S, Yamamoto S, Nakajima N, Sakakibara S, Watanabe K, Inagaki S, Nagayama Y, Narushima Y, Yamada H {\em et~al.\/} 2004 {\em Nuclear fusion\/} {\bf 44} 217

\bibitem{tikhonchuk2018physics}
Tikhonchuk V 2018 {\em Nuclear Fusion\/} {\bf 59} 032001

\bibitem{frenje2020nuclear}
Frenje J~A 2020 {\em Plasma Physics and Controlled Fusion\/} {\bf 62} 023001

\bibitem{hundhausen1970composition}
Hundhausen A 1970 {\em Reviews of Geophysics\/} {\bf 8} 729--811

\bibitem{treumann1997advanced}
Treumann R~A and Baumjohann W 1997 {\em Advanced space plasma physics\/} vol~30 (Imperial College Press London)

\bibitem{koepke2008interrelated}
Koepke M~E 2008 {\em Reviews of geophysics\/} {\bf 46}

\bibitem{tsurutani2022space}
Tsurutani B~T, Zank G~P, Sterken V~J, Shibata K, Nagai T, Mannucci A~J, Malaspina D~M, Lakhina G~S, Kanekal S~G, Hosokawa K {\em et~al.\/} 2022 {\em IEEE Transactions on Plasma Science\/} {\bf 51} 1595--1655

\bibitem{kortshagen2016nonthermal}
Kortshagen U~R, Sankaran R~M, Pereira R~N, Girshick S~L, Wu J~J and Aydil E~S 2016 {\em Chemical reviews\/} {\bf 116} 11061--11127

\bibitem{ravi2009coagulation}
Ravi L and Girshick S~L 2009 {\em Physical Review E—Statistical, Nonlinear, and Soft Matter Physics\/} {\bf 79} 026408

\bibitem{donders2022characterization}
Donders T, Staps T and Beckers J 2022 {\em Journal of Physics D: Applied Physics\/} {\bf 55} 395203

\bibitem{ramkorun2024introducing}
Ramkorun B, Jain S, Taba A, Mahjouri-Samani M, Miller M~E, Thakur S~C, Thomas E and Comes R~B 2024 {\em Applied Physics Letters\/} {\bf 124}

\bibitem{van2015fast}
Van~de Wetering F, Brooimans R, Nijdam S, Beckers J and Kroesen G 2015 {\em Journal of Physics D: Applied Physics\/} {\bf 48} 035204

\bibitem{greiner2011nanodust}
Greiner F, Carstensen J, Ketelsen H and Piel A 2011 Nanodust in magnetized plasma {\em AIP Conference Proceedings\/} vol 1397 (AIP Publishing) pp 315--316

\bibitem{merlino2007understanding}
Merlino R~L 2007 {\em American Journal of Physics\/} {\bf 75} 1078--1085

\bibitem{thomas2010driven}
Thomas E 2010 {\em Physics of Plasmas\/} {\bf 17}

\bibitem{melzer2020dust}
Melzer A, Kr{\"u}ger H, Sch{\"u}tt S and Mulsow M 2020 {\em Physics of Plasmas\/} {\bf 27} 033704

\bibitem{villaflor2026detecting}
Villaflor J, Hyden H, Pizzutti C, Buckner A, Sisolekar S and Ramkorun B 2026 Detecting changes in optical emission spectroscopy intensity in a magnetized nonthermal plasma to grow nanoparticles {\em 2026 IEEE International Conference on Plasma Science (ICOPS)\/} (IEEE) pp 267--267

\bibitem{buckner2026optical}
Buckner A, Villaflor J, Hyden H, Sisolekar S, Pizzutti C and Ramkorun B 2026 Optical emission spectroscopy analysis of a magnetized argon glow discharge plasma to grow nanoparticles {\em 2026 IEEE International Conference on Plasma Science (ICOPS)\/} (IEEE) pp 259--259

\bibitem{hyden2026studying}
Hyden H, Pizzutti C, Buckner A, Villaflor J, Sisolekar S and Ramkorun B 2026 Studying glow discharge argon nonthermal plasma electron temperature and density to grow nanoparticles {\em 2026 IEEE International Conference on Plasma Science (ICOPS)\/} (IEEE) pp 260--260

\bibitem{lewandowski2014colorado}
Lewandowski H 2014 Colorado learning about science survey for experimental physics (e-class) {\em APS March Meeting Abstracts\/} vol 2014 pp S38--003

\bibitem{wilcox2016students}
Wilcox B~R and Lewandowski H~J 2016 {\em Physical Review Physics Education Research\/} {\bf 12} 010123

\bibitem{werth2023enhancing}
Werth A, West C~G, Sulaiman N and Lewandowski H 2023 {\em Physical Review Physics Education Research\/} {\bf 19} 020151

\bibitem{liu2025students}
Liu Q and Lewandowski H 2025 {\em Physical Review Physics Education Research\/} {\bf 21} 020151

\end{thebibliography}
\end{document}